\documentclass[10pt,conference]{IEEEtran}

\usepackage{amsmath}
\usepackage{amssymb}
\usepackage{verbatim}
\usepackage{epsfig}
\usepackage{subfigure}
\usepackage{pstricks}
\usepackage{pstricks-add}
\usepackage{bbm}

\begin{document}
\title{Opportunistic Routing in Ad Hoc Networks:\\How many relays should there be?\\What rate should nodes use?}
\author{
\authorblockN{Joseph Blomer, Nihar Jindal}
\authorblockA{Dept. ECE, University of Minnesota, Minneapolis, MN, USA 55455, 
\\ Email: {\tt \{blome015,nihar\}@umn.edu}\\
}
}\maketitle
\begin{abstract}
Opportunistic routing is a multi-hop routing scheme which allows for selection of the best immediately available relay.  In blind opportunistic routing protocols, where transmitters blindly broadcast without knowledge of the surrounding nodes, two fundamental design parameters are the node transmission probability and the transmission spectral efficiency. In this paper these parameters are selected to maximize end-to-end performance, characterized by the product of transmitter density, hop distance and rate.  Due to the intractability of the problem as stated, an approximation function is examined which proves reasonably accurate. Our results show how the above design parameters should be selected based on inherent system parameters such as the path loss exponent and the noise level.
\end{abstract}
\section{Introduction}

Ad hoc networks operate on the basis of multi-hop routing, which allows information to be communicated across the network over a series of hops.
End-to-end performance depends critically on the quality of the multi-hop routes used, but choosing good routes in a dynamic network (e.g. one where nodes are moving quickly) is a particularly difficult task.
Opportunistic routing (e.g \cite{Baccelli06} - \cite{ZorziPupolin95}) is well suited to such dynamic settings, as it can be performed with low overhead while also exploiting spatial diversity (in terms of fading and topology).

In this paper we focus on a purely opportunistic routing protocol, in which nodes broadcast packets in a completely blind manner (i.e. without any knowledge of surrounding nodes).
More specifically, the protocol is described as follows:
\begin{enumerate}
\item In each time slot, each node randomly decides, with some fixed probability, to either transmit or receive.
\item A transmitting node will blindly broadcast a packet at some fixed spectral efficiency, whereas receivers listen for any transmission.
\item All receiver nodes that successfully decode the packet will send an ACK back to the transmitting node during an acknowledgement period.  The ACK packets contain absolute geographic information about the RX location.
\item The transmitter selects the successful receiver which offers the most progress towards the packet's final destination and sends a message to that node electing it to become the next forwarder.
\end{enumerate}

This protocol allows for the instantaneous choice of the best next hop, at the cost of an acknowledgement period and with the requirement for geographical information.
Within this algorithm, the key adjustable parameters are the \textit{transmission probability} and the transmission \textit{spectral efficiency}.

The transmission probability, denoted by $p$, determines the proportion of transmitters to receivers in each slot. 
When $p$ is large, there is a large amount of interference between the nodes.
This interference will cause only receivers close to transmitters to be potential forwarders.  
When the transmission probability is low, there are fewer simultaneous transmissions and thus less interference and more available relays.  
Thus longer hops are possible, but in fewer numbers. 
As a result, the trade-off with transmission probability is essentially between many simultaneous short hops, or fewer long hops.

When the spectral efficiency is high, a large signal-to-interference ratio (SIR) is required to decode.  
Thus, only relays which are close to the transmitter are likely to decode.
On the other hand, a lower spectral efficiency allows nodes that are farther away to decode.
Therefore, the trade-off with spectral efficiency is between shorter hops at higher data rate or longer hops at a lower data rate.

The objective of this paper is determining the transmission probability and spectral efficiency that optimally balance these trade-offs.
We study blind opportunistic routing in a spatial model, as in \cite{Baccelli06}.  
In this model, end-to-end performance is characterized by the forward-rate-density, which is the product of average transmitter density, average hop distance and transmission rate (spectral efficiency). 
This is a function of measured network properties (e.g. path-loss exponent) and adjustable parameters (spectral efficiency and transmission probability).  
Our objective is to find the transmission probability and transmission spectral efficiency that maximize this metric.
Studying this quantity directly is intractable, however we derive a reasonably accurate approximation and reinforce the conclusions with Monte Carlo simulations.

\section{Preliminaries}

\subsection{Network Model}
\label{sec:netmodel}

Consider an infinite set of transmit/receive nodes $\Phi$ distributed according to a homogeneous 2-D Poisson point process (PPP) with density $\lambda$ $\text{m}^{-2}$. 
We consider a slotted transmission scheme. In each slot a node elects to become a transmitter with probability $p$, independent across users and slots.
The set of locations of the transmitters (TX1, TX2, ..) denoted $\Phi^t$ and the set of locations of the receivers $\Phi^r$ then form independent PPP's of intensity $\lambda p$  and $\lambda (1-p)$ respectively. 
\subsection{Channel Model}
We consider a path-loss model with exponent $\alpha > 2$ and Rayleigh fading coefficients $h_{i,j}$ from TX $i$ to RX $j$.  Denoting the signal transmitted by TX$i$ (located at $X_i$) as $u_i$, the received signal of RX$j$ (at location $X_j$) is given by:

\begin{eqnarray}
Y_j & = & \sum_{i} h_{i,j} |X_i-X_j|^{-\alpha/2} u_i + z.
\end{eqnarray}

For the time being, consider the network is interference limited, so thermal noise ($z$) is negligible. The case of a channel with thermal noise is considered in Sec \ref{sec:numerical_results}. 

Assuming the transmit powers are all equal, the SIR (signal to interference ratio) from TX $i$ to RX $j$ is defined:
\begin{eqnarray}
\label{eqn:sir}
S_{i,j} \triangleq \frac{|h_{i,j}|^2 |X_i-X_j|^{-\alpha} }{ \sum_{k \neq i} |h_{k,j}|^2 |X_k -X_j|^{-\alpha}}.
\end{eqnarray}
We assume that all users transmit at rate equal to $R(\beta) = \log_2(1 + \beta)$; thus a communication is successful if and only if the received SIR is larger than a threshold value $\beta$.

\section{Progress-Rate Product}
Although we are interested in the network-wide performance of the protocol, by the homogeneity of the PPP this is statistically equivalent to the performance of any transmitter.
Thus, without loss of generality we can focus on TX$0$ located at the origin.
From the perspective of TX$0$, all the other TX nodes form a homogeneous PPP with intensity $\lambda p$ (by Slivnyaks Theorem \cite{Stoyan96}, the distribution is unaffected by conditioning on the presence of TX$0$).
The transmitters and receivers are ordered by their distance to the origin i.e. TX1 is closest to the origin, TX2 the next, etc.
For the sake of simplicity, the SIR from TX$0$ to RX$j$ is denoted $S_j$ and is given by $S_{j} = S_{0,j}$.
\begin{figure}[t]
	\centering
	\includegraphics[width=3.25in]{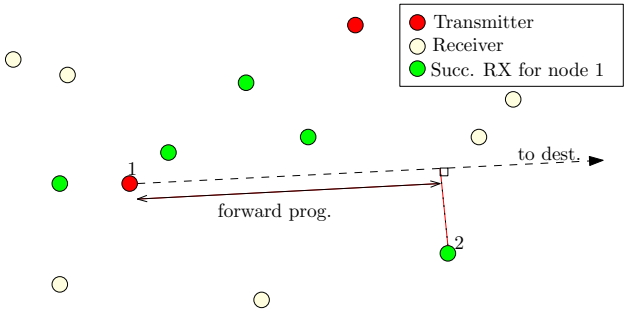}
	\caption{Example network: TX node 1 has a message to be sent along dashed line.  Among successful message receivers, RX node 2 offers the most forward progress and will forward the packet in the next slot.}
	\label{fig:net_ex}
\end{figure}
\subsection{Forward progress density}
\label{sec:forprog}
We assume that the transmitter located at the origin has a packet to be sent to a receiver a very large distance away (e.g., fig. \ref{fig:net_ex}).  
Specifically, we assume the final destination is located an infinite distance away along the $x$-axis.  
Thus, the relay offering the most progress is the one with the maximum $x$-coordinate amongst the set of receivers that successfully decode TX$0$'s packet.  The corresponding progress, for a given network realization, is: 
\begin{eqnarray}
&& \nonumber D(\Phi^r,\Phi^t,\beta) \triangleq \\
 && \, \max_{X_j \in \Phi^r} \Big[ \mathbf{1}(S_j \geq \beta | \Phi^t) \Big( |X_j| \cdot \cos\big(\theta \left( X_j \right) \big) \Big) \Big] 
\end{eqnarray}
where $\mathbf{1}(S_j \geq \beta | \Phi^t)$ is the indicator function that the SIR at RX$j$ (at point $X_j$) relative to TX$0$ (at the origin) is greater than the threshold.
The expectation of this function with respect to the node process and fading is $d(p,\beta)$.
\begin{eqnarray}
d(p,\beta) & \triangleq & \mathbb{E}^{\Phi,|h|^2}\left[ D(\Phi^r,\Phi^t,\beta) \right]
\end{eqnarray}
This quantity is the expected progress (towards the destination) made in each hop.
The forward progress density $f(p,\beta)$ then is:
\begin{eqnarray}
\label{eqn:mfp}
f(p,\beta) \triangleq d(p,\beta)\lambda p.
\end{eqnarray}
Finally, the forward progress-rate-density function $K(p,\beta)$ is 
\begin{eqnarray}
\nonumber
K(p,\beta) & \triangleq & \left(\lambda p\right) \cdot \mathbb{E}[D(\Phi^r,\Phi^t,\beta)] \cdot \log_2(1+\beta)\\
& = & f(p,\beta) \cdot R(\beta).
\end{eqnarray}

Assuming each node in the network wishes to send data to another node a distance $L$ away, at most $p \cdot d(p,\beta)/L$ packets can originate at each node in each slot \cite{Baccelli06}.
Translating to bits/sec/Hz, this means that the maximum end-to-end data rate (per node) is 
\begin{equation}
p \cdot d(p,\beta) \cdot \log_2 (1+\beta)/L \text{ [bps/Hz]}.
\end{equation}
Thus, $K(p,\beta)$ is directly proportional to end-to-end rate.  

The quantity $K(p,\beta)$, which is deterministic, is a function of transmission probability $p$ and SIR threshold $\beta$.  We seek the optimal values of $\beta$ and $p$ which maximize this product:
\begin{equation}
(p^*,\beta^*) = \arg\max_{p,\beta} K(p,\beta).
\end{equation}
A closed-form expression for $K(p,\beta)$ cannot be found in general, so we rely on an approximation which is developed in the following section.

\subsection{SIR Cell-Based Approximation}
\label{sec:sircell}
The basis of our approximation to $K(p,\beta)$ is the concept of an SIR cell \cite{Baccelli09V1}, \cite{Baccelli09V2}.
In the absence of fading, any RX within a particular region (i.e., the SIR cell) around a TX can decode, and the best forwarder is the RX in this region with the largest $x$-coordinate.
Therefore, the forward progress is determined by the SIR cell -- which is completely determined by the interferer locations and thus is a function only of the interferer process -- and by the receiver locations (i.e., $\Phi^r$).
To reach a tractable approximation for $K(p,\beta)$, we remove the randomness in the SIR cell and assume it is a deterministic region that is a function of $p$ and $\beta$, while retaining the randomness in $\Phi^r$.

We consider the average size of the SIR cell.  This quantity is clearly defined in the absence of fading.  
On the other hand, the SIR cell is not well defined with fading because the SIR depends on independent fading RV's (specific to each receiver location) in addition to the interferer locations.
However, we can derive a quantity analogous to the SIR cell area, denoted by $v_0$, by integrating the point-wise success probability over all space.  From \cite{Baccelli09V2}, the success probability for a RX a distance $y$ from the TX is:
\begin{eqnarray}
\nonumber p_0(y) & = & \mathbb{P}^{\Phi^t}[S_j \geq \beta ] \\
& = & \exp \left(\dfrac{-\pi \lambda p |y|^2 \beta^{\frac{2}{\alpha}}}{G(\alpha)}\right).
\end{eqnarray}

Thus, $v_0$ is given by:
\begin{eqnarray}
  v_0  & = & \int_{y \in \mathbb{R}^2} p_0(y) dy  = \dfrac{G(\alpha)}{\lambda p \beta^{\frac{2}{\alpha}}}.  \label{eqn:sir_cell_volume}
\end{eqnarray}
where
\begin{equation}
G(\alpha) = \frac{\alpha}{2 \Gamma(\frac{2}{\alpha})\Gamma(1-\frac{2}{\alpha})}
\end{equation}
and $\Gamma(z)$ is the gamma function $\Gamma(z) = \int_0^\infty t^{z-1}e^{-t} dz$.

\begin{figure}
	\centering
	\includegraphics[width=2in]{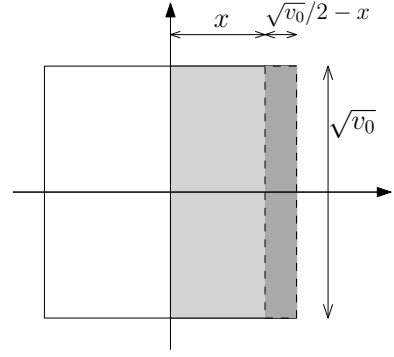}
	\caption{SIR cell is assumed to be a square $V_0$ of side length $\sqrt{v_0}$. The probability that the forward progress is $x$ is the probability there are no RX in the dark shaded area.}
	\label{fig:method2}
\end{figure}

We will then assume the SIR cell has area equal to its mean.  
We now approximate the SIR cell by a square centered on TX$0$ with side $\sqrt{v_0}$.
In other words, we assume that all RX within this square are able to decode (ref. fig. \ref{fig:method2}).
$V_0^+$ is the positive half of that square with area $v_0/2$.
Under these assumptions, the expected forward progress is:

\begin{eqnarray}
\nonumber && \tilde{d}_{sq} (p,\beta) \\
\nonumber & = & \mathbb{E}\left[\max_{X_j \in V_0^+} x_j \right]\\
\nonumber & = & \mathbb{E}\left[\mathbb{E}\left[\max_{X_j \in V_0^+} x_j \Big\vert \#\left( \{X_j \in V_0^+\} \right) = j\right]\right] \\
\nonumber & = & \sum_{j=0}^\infty \mathbb{E}\left[ \max_{X_i \in V_0^+} x_i \Big\vert \#\left( \left\{X_i \in V_0^+ \right\}\right) = j \right] \times \\
\nonumber && \left(  e^{ -\lambda \cdot (1-p) \cdot \frac{v_0}{2}}\dfrac{(\lambda \cdot (1-p) \cdot \frac{v_0}{2})^j}{j!} \right)\\
\nonumber &=& \sum_{j=0}^\infty \left( \dfrac{\sqrt{v_0}}{2} \dfrac{j}{j+1} \cdot  e^{ -\lambda \cdot (1-p) \cdot \frac{v_0}{2}}\dfrac{(\lambda \cdot (1-p) \cdot \frac{v_0}{2})^j}{j!} \right)\\
&=& \dfrac{\sqrt{v_0}}{2} \left( \frac{e^{-c}-(1-c)}{c} \right), \label{eqn:fp}
\end{eqnarray}
where $\#(\cdot)$ gives the cardinality of a set and
\begin{eqnarray}
c & = & \dfrac{ \lambda\cdot (1-p) \cdot v_0 }{2} = \dfrac{(1-p)}{p}\frac{ G(\alpha)}{ 2 \beta^{2/\alpha} }.
\end{eqnarray}
Note that the value $c$ is the expected number of receivers in the half-square $V_0^+$; it is also equal to half the average number of successful outgoing transmissions from a TX as in \cite{WeberJindal08}.

Thus, our approximation to the progress-rate-density is:
\begin{eqnarray}
\label{eqn:mfp_eqn}
f(p,\beta) & \approx & \tilde{f}_{sq}(p,\beta) = \tilde{d}_{sq}(p,\beta) (\lambda p)
\\
\label{eqn:MFR_eqn}
K(p,\beta) & \approx & \tilde{K}_{sq}(p,\beta)= \tilde{f}_{sq}(p,\beta) \log_2(1+\beta).
\end{eqnarray}
The optimal values then are the solution to the equation
\begin{equation}
\dfrac{\partial}{\partial p} \dfrac{\partial}{\partial \beta} \tilde{K}_{sq}(p,\beta) = 0.
\end{equation}
Although there is no closed form solution to this equation, the approximation nonetheless yields valuable insight. 
By rearranging the terms in (\ref{eqn:fp}), we can interpret the forward progress approximation as the product of the maximum possible relay distance (i.e., half the side-length of the SIR cell) and a fractional term that captures how much of the maximum is attained by the best receiver.
\begin{eqnarray}
\tilde{d}_{sq} (p,\beta) & = & \underbrace{\dfrac{\sqrt{v_0}}{2}}_\text{max. relay dist.} \underbrace{\left( 1- \frac{1-e^{-c}}{c} \right)}_\text{frac. of max.}
\end{eqnarray}
The value $v_0$ is decreasing in $p$: as $p$ increases, the amount of interference between transmitters is increasing, giving each transmitter a smaller expected SIR cell size.  The value of $c$ is also decreasing in $p$ as a result of decreased $v_0$ and smaller receiver density.

The value $\tilde{K}_{sq}(p,\beta)$ can be written:
\begin{eqnarray}
\nonumber & &\tilde{K}_{sq}(p,\beta)\\
& = & \frac{1}{2} \log_2 (1+\beta) \sqrt{\lambda p G(\alpha) \beta^{\frac{-2}{\alpha} }}\left( 1- \frac{1-e^{-c}}{c}\right)
\end{eqnarray}
which lends the intuition that the end-to-end performance is increasing with $p$ until the $1- \frac{1-e^{-c}}{c}$ fractional term becomes too small.
When the value of $c$ (expected potential forwarders) is greater than 3, this fractional term is greater than $2/3$. When $p$ is large enough to cause $c$ to go below 3, the performance drops off sharply.
\section{Numerical Results}
\label{sec:numerical_results}
For numerical results, Monte-Carlo simulations were performed.  Locations of receivers and transmitters were realized using the properties of the PPP (c.f. \cite{Haenggi05}).  SIR was calculated at each receiver and the largest progress offered was averaged.

\subsection{Accuracy, Path-loss and Fading}
\begin{figure}
	\centering
	\includegraphics[width=3in]{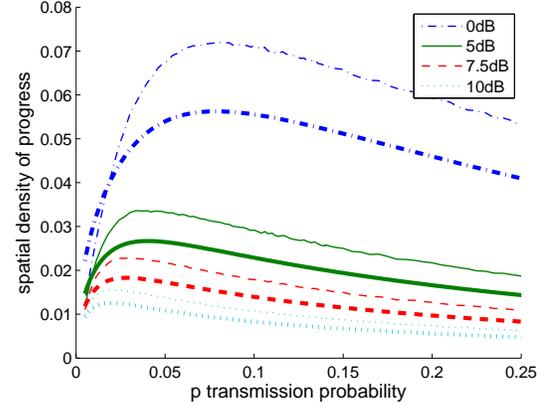}
	\caption{Spatial density of progress: approximation (thick curves) and simulation for $\lambda=1$, $\alpha=3$, varying SIR threshold $\beta$.}
	\label{fig:mfp_vs_p}
\end{figure}

Fig. \ref{fig:mfp_vs_p} shows the accuracy of the approximation of the spatial density of progress (\ref{eqn:mfp}), (\ref{eqn:mfp_eqn}) for $p$ and $\beta$.  While the approximation is not absolutely accurate, it does correctly capture the dependence on $p$ and $\beta$.

Fig. \ref{fig:arg_max} shows the best $p$ or $\beta$ maximizing the progress-rate-density (\ref{eqn:MFR_eqn}) for a fixed $\beta$ or $p$, respectively.  Again, the approximation helps to answer the essential question of what parameter choices are optimal.
\footnote{ Simulation results are similar for a non-fading environment. }

\begin{figure}
	\centering
	\includegraphics[width=3in]{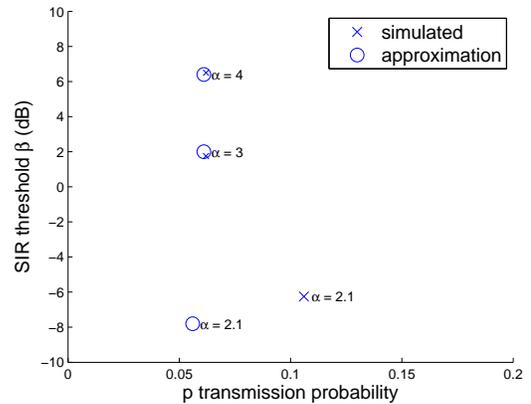}
	\caption{Optimal choice of $p$ and $\beta$ giving maximum progress-rate for fixed $\alpha$ path-loss values.}
	\label{fig:arg_max_over_a}
\end{figure}

In Fig. \ref{fig:arg_max_over_a} the jointly optimal $(p,\beta)$ are plotted for different values of the path loss exponent.
Note that the optimal value of $p$ is essentially the same for $\alpha=3$ and $4$, but that the optimizing SIR threshold increases with the path loss exponent.  
This increase is similar to \cite{JindalAndrews08}, where the optimizing SIR threshold is found for a slightly different, but related, problem.

\subsection{Sensitivity to parameter variation}
\label{sec:sens_par}
\begin{figure}
	\centering
	\includegraphics[width=3in]{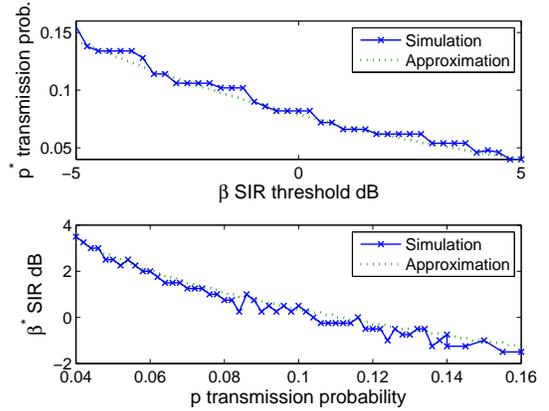}
	\caption{Optimal $p$ values giving best progress-rate-density for fixed values of SIR $\beta$ (top).
	Optimal $\beta$ values giving maximum progress-rate for fixed values $p$ (bottom).}
	\label{fig:arg_max}
\end{figure}

\begin{figure}
	\centering
	\includegraphics[width=3in]{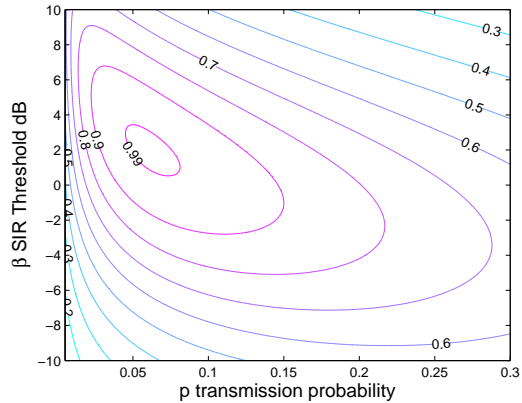}
	\caption{Contour plot of normalized forward-rate-density function approximation.}
	\label{fig:approx_contour}
\end{figure}

Fig. \ref{fig:approx_contour} shows the normalized approximation of Eqn. (\ref{eqn:MFR_eqn}) for path loss $\alpha=3$ defined by 
\begin{equation}
\tilde{K}_{\text{sq,norm}}(p,\beta) \triangleq \frac{\tilde{K}_{sq}(p,\beta) }{  \max_{(p,\beta)}\tilde{K}_{sq}(p,\beta) }.
\end{equation}
This figure shows how the performance decreases with different values of $p$ and $\beta$.
Observe that there exist a wide range of $(p,\beta)$ pairs that are near optimal (within 90\%).  
If there exists a strong reason to choose a particular $\beta$ (e.g., limited code rates \& modulations), or a particular $p$ (e.g., power cycling, energy saving from different TX and RX powers), end-to-end performance does not suffer as long as the other parameter is appropriately chosen (to figs. \ref{fig:arg_max},  \ref{fig:approx_contour})

\subsection{Robustness to Noise}
\begin{figure}
	\centering
	\includegraphics[width=3in]{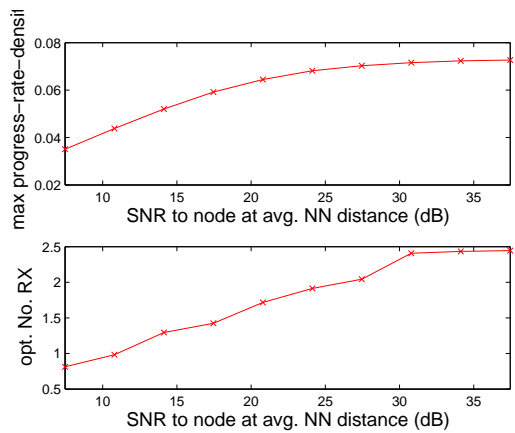}
	\caption{Simulated best progress-rate-density achievable for varying SNR to node at mean nearest-neighbor distance
	$(\lambda \pi)^{\alpha/2}/\sigma^2$}
	\label{fig:max_pr_vs_noise}
\end{figure}

In order to simulate the effect of noise on the forward progress, we modify the SIR (\ref{eqn:sir}) to include noise power $\sigma^2$.
\begin{eqnarray}
S_{i,j} = \frac{|h_{i,j}|^2 |X_i-X_j|^{-\alpha} }{ \sum_{k \neq i} |h_{k,j}|^2 |X_k -X_j|^{-\alpha}+\sigma^2}
\end{eqnarray}
The metric for comparison is the average SNR to a node located at the average nearest-neighbor distance ($d_{NN}$).
\begin{eqnarray}
\overline{SNR}_{NN} & = & \frac{ \mathbb{E}\left[ d^2_{NN} \right]^{-\frac{\alpha}{2} }  }{\sigma^2} = \frac{(\lambda \pi)^{\alpha/2}}{\sigma^2}
\end{eqnarray}
Fig. \ref{fig:max_pr_vs_noise} shows the maximum forward-rate-density is decreasing with increasing channel noise. Also, the number of potential forwarders is decreasing with increased channel noise.

\begin{figure}
	\centering
	\includegraphics[width=3in]{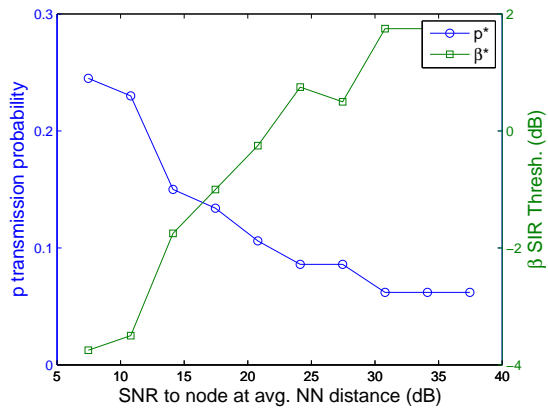}
	\caption{Optimal values of $p$, $\beta$ for varying SNR to node at mean nearest-neighbor distance
	$(\lambda \pi)^{\alpha/2}/\sigma^2$}
	\label{fig:p_b_star_vs_noise}
\end{figure}

Fig. \ref{fig:p_b_star_vs_noise} shows the effects of channel noise power on optimal $p$ and $\beta$.  The value of $p^*$ is increasing and the value of $\beta^*$ is decreasing with higher noise floor.
The maximum possible spectral efficiency is expected to drop with increasing noise power.
As explored in \ref{sec:sens_par}, when $\beta$ is fixed to some value (in this case as a result of the noise floor), the transmission probability can increase to compensate.
\section{Conclusion}
We studied the case of opportunistic routing in an ad hoc wireless network with the goal of maximizing the product of forward progress density and the rate of data transmission.  We developed an approximation using the concept of SIR cells and found for $\alpha=3$ the optimal spectral efficiency is near $1.3$ bps/Hz and the optimal probability of transmission is near $0.06$. The results give the intuition that having a low transmission probability ensures a low amount of interference and a relatively high SIR threshold ensures only local transmissions are received.


\end{document}